\documentclass[pra,twocolumn,floatfix,a4paper,superscriptaddress]{revtex4}

\usepackage{bm,color,amsmath,txfonts}

\usepackage{graphicx}
\usepackage{subfigure}
\usepackage{verbatim}
\usepackage{dcolumn}
\usepackage{bm}
\usepackage{epsf}
\usepackage{xcolor}
\usepackage{hyperref}
\usepackage{hhline}
\usepackage{float}
\usepackage{enumerate}
\usepackage{bbm}
\usepackage{lipsum}
\definecolor{mygreen}{rgb}{0.01, 0.31, 0.59}
\definecolor{myblue}{rgb}{0.01, 0.31, 0.59}
\hypersetup{
 colorlinks=true,   
 linkcolor=blue,  
 citecolor=blue, 
 urlcolor =blue    
}

\hypersetup{
 colorlinks=true,   
 linkcolor=blue,  
 citecolor=blue, 
 urlcolor =blue    
}

\newcommand{\CC}{{\cal C}}

\newcommand{\MM}{{\cal M}}

\begin{document}

\title{Optical sensing of magnons via the magnetoelastic displacement}

\author{Zhi-Yuan Fan}
\affiliation{Interdisciplinary Center of Quantum Information, State Key Laboratory of Modern Optical Instrumentation, and Zhejiang Province Key Laboratory of Quantum Technology and Device, Department of Physics, Zhejiang University, Hangzhou 310027, China}
\author{Rui-Chang Shen}
\affiliation{Interdisciplinary Center of Quantum Information, State Key Laboratory of Modern Optical Instrumentation, and Zhejiang Province Key Laboratory of Quantum Technology and Device, Department of Physics, Zhejiang University, Hangzhou 310027, China}
\author{Yi-Pu Wang}
\affiliation{Interdisciplinary Center of Quantum Information, State Key Laboratory of Modern Optical Instrumentation, and Zhejiang Province Key Laboratory of Quantum Technology and Device, Department of Physics, Zhejiang University, Hangzhou 310027, China}
\author{Jie Li}\thanks{jieli007@zju.edu.cn}
\affiliation{Interdisciplinary Center of Quantum Information, State Key Laboratory of Modern Optical Instrumentation, and Zhejiang Province Key Laboratory of Quantum Technology and Device, Department of Physics, Zhejiang University, Hangzhou 310027, China}
\author{J. Q. You}
\affiliation{Interdisciplinary Center of Quantum Information, State Key Laboratory of Modern Optical Instrumentation, and Zhejiang Province Key Laboratory of Quantum Technology and Device, Department of Physics, Zhejiang University, Hangzhou 310027, China}

\begin{abstract}
We show how to measure a steady-state magnon population in a magnetostatic mode of a ferromagnet or ferrimagnet, such as yttrium iron garnet. We adopt an optomechanical approach and utilize the magnetoelasticity of the ferromagnet. The magnetostrictive force dispersively couples magnons to the deformation displacement of the ferromagnet, which is proportional to the magnon population. By further coupling the mechanical displacement to an optical cavity that is resonantly driven by a weak laser, the magnetostrictively induced displacement can be sensed by measuring the phase quadrature of the optical field. The phase shows an excellent linear dependence on the magnon population for a not very large population, and can thus be used as a `magnometer' to measure the magnon population. We further study the effect of thermal noises, and find a high signal-to-noise ratio even at room temperature. At cryogenic temperatures, the resolution of magnon excitation numbers is essentially limited by the vacuum fluctuations of the phase, which can be significantly improved by using a squeezed light.
\end{abstract}

\date{\today}
\maketitle

\section{Introduction}

Ever since the successful demonstration of the strong coupling between microwave cavity photons and ferrimagnetic magnons in yttrium iron garnet (YIG)~\cite{S1,S2,S3}, the field of cavity magnonics has attracted considerable attention and achieved significant progress~\cite{NakaRev,NatRev,RMP}. Such a strongly coupled system has become a new platform for studying rich and stimulating phenomena belonging to cavity quantum electrodynamics. This benefits largely from the distinct advantages of YIG, such as the high spin density, low damping rate, and rich nonlinear interactions between excitations in the crystal.

Many studies in cavity magnonics require the estimation of the magnon population in a magnetostatic mode of a ferromagnet or ferrimagnet. For example, it can be used 
to determine the effective {\it dispersive} coupling strength in either the magnon-qubit interaction~\cite{Nak17,Nak20S} or the magnon-phonon interaction~\cite{JL18,JL19}. The method of measuring the transmission (or reflection) spectrum of the microwave cavity in the conventional cavity-magnon system~\cite{S1,S2,S3} is not able to infer the magnon population because their linear beamsplitter coupling is independent of the magnon population. One can, however, implement state tomography of microwave photons that have a beamsplitter interaction with magnons. In this way, the magnon population can be determined, since the population is just the square of the amplitude in the phase space. The tomography of magnonic states has been realized in Ref.~\cite{prb21}.  There are approaches for measuring a large magnon population based on, e.g., electromagnetic induction~\cite{Melkov}, the inverse spin-Hall effect~\cite{invSH1,invSH2,invSH3,prb21}, Brillouin light scattering~\cite{BLS1,BLS2}, etc. Alternatively, one may consider utilizing the magnon Kerr nonlinearity, where the Kerr effect leads to a frequency shift of the magnon mode dependent on the magnon population~\cite{YP16,YP18,YP21}. Thus the magnon population can be inferred by measuring the frequency shift. This approach works also for a large magnon population (under a strong pump). For a small magnon population, it can be realized by {\it dispersively} coupling magnons to a superconducting qubit~\cite{Nak17,Nak20S,Nak20L}, where either the qubit frequency~\cite{Nak17} or dephasing~\cite{Nak20L} depends on the magnon population. Therefore, by measuring the frequency shift or coherence time of the qubit, the magnon excitations can be resolved. These approaches can achieve very high resolutions of magnons, but typically work at very low temperatures.  

Here, we introduce a new approach by adopting the magnetostrictive (magnetoelastic) coupling between magnons and the vibrational motion of a ferrimagnet, e.g. YIG. The magnetostrictive force leads to the geometric deformation of the ferrimagnet, forming vibrational modes (phonons)~\cite{Kittle}. We consider the situation where the phonon frequency is much smaller than the magnon frequency~\cite{JL18,Tang,Davis}, such that their dispersive interaction becomes dominant, leading to the deformation displacement proportional to the magnon population. This is analogous to the displacement of a mechanical oscillator 
induced by the radiation pressure in optomechanics~\cite{OMrmp}. Therefore, by measuring the displacement one can infer the magnon population. To do it, we adopt an optomechanical approach by coupling the displacement to an optical cavity. When the cavity is {\it resonantly} driven, a high-sensitivity measurement of the mechanical displacement can be realized by measuring the phase of the cavity field. To minimize the disturbance to the displacement due to radiation pressure, we use a weak probe light. This guarantees that the frequency shift of the optical cavity is mainly caused by the magnomechanically induced displacement. For a not very large magnon population, we find an excellent linear dependence of the optical phase on the magnon population. 
Therefore, in this linear regime the optical phase can act as a meter for the magnon population. We further study the effect of thermal noises on our protocol, and find that the signal-to-noise ratio (SNR) can be very high even at room temperature. The resolution of magnons is essentially limited by the vacuum fluctuations of the phase when working at cryogenic temperatures, e.g., tens of millikelvin. The resolution can be significantly improved by feeding a squeezed light into the cavity, which can suppress the phase noise to be well below the standard quantum limit.

This article is organized as follows. In Sec.~\ref{sys}, we introduce our model and provide its Hamiltonian and quantum Langevin equations. We then solve the equations and obtain steady-state solutions for the classical averages, which lead to the main result of the work: the magnon population can be measured by the optical phase based on their linear dependence. In Sec.~\ref{noise}, we study the effect of thermal noises on our protocol, which lead to phase fluctuations and a finite SNR. We derive the analytical expression of the phase noise and show that a high SNR can still be achieved in the presence of thermal noises, even at room temperature. We discuss how to improve the SNR and the resolution of magnons. Finally, we summarize in Sec.~\ref{conc}.

\begin{figure}
\includegraphics[width=\linewidth]{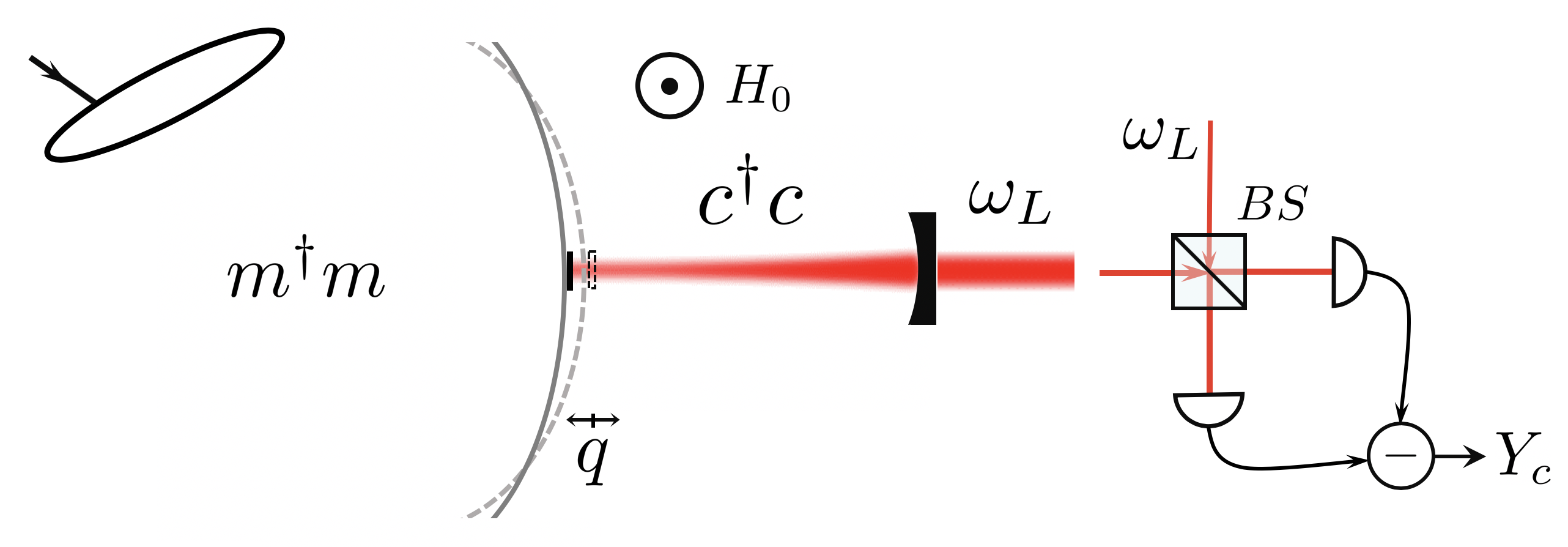}
\caption{Sketch of the protocol for optically measuring magnon population in a ferromagnet. By using a homodyne setup, the magnetostrictively induced displacement, proportional to the magnon population, is read out in the phase quadrature of the optical field. BS: beamsplitter. See the text for a detailed description. } 
\label{fig1}
\end{figure}

\section{The protocol}
\label{sys}

We consider a magnomechanical system, e.g., in a YIG sphere~\cite{note}, which consists of a magnon mode (e.g., the uniform-precession Kittel mode~\cite{Kittel}), and a phonon mode of deformation vibration, as sketched in Fig.~\ref{fig1}. The magnon mode represents the collective motion (spin wave) of a large number of spins in the ferrimagnet, and is activated by placing the ferrimagnet in a uniform bias magnetic field and applying, e.g., via a loop coil~\cite{Naka16}, a microwave drive field with its magnetic component perpendicular to the bias field.  The magnon mode couples to a deformation phonon mode of the ferrimagnet via the magnetostrictive interaction. We consider the situation where the frequency of the phonon mode is much lower than that of the magnon mode~\cite{JL18,Tang,Davis}, such that they couple via a radiation pressure-like dispersive interaction. Such a nonlinear coupling has been recognized as a cornerstone of many quantum protocols~\cite{JieNJP, Tan, Kong, Ding, JL20, Davis20, Yang, QST, NSR, HFW, TB,Jing, JHW, HJ, CLi}. Relevant experiments have demonstrated magnomechanically induced transparency/absorption~\cite{Tang} and mechanical cooling/lasing~\cite{Davis}. 

We aim to probe the magnons in the ferrimagnet, and particularly measure its steady-state population. To this end, we adopt an optomechanical approach by coupling the magnetostrictively induced displacement to an optical cavity. This can be realized, e.g., by attaching a high-reflectivity mirror pad~\cite{SG} to the surface of the ferrimagnet. The mirror is so small that it does not appreciably affect the mechanical properties of the ferrimagnet. Therefore, through this way the magnomechanical displacement can be probed by light exploiting the optomechanical interaction, which is then used to determine the magnon population. 
The corresponding Hamiltonian of the magnomechanical system combined with the optical cavity reads
\begin{equation}
H= H_0 + H_{\rm int} + H_{d},
\end{equation}
where 
\begin{equation}
\begin{split}
H_0/\hbar=&\, \omega_m m^\dagger m+\frac{\omega_b}{2}\left ( q^2 + p^2  \right ) + \omega_c c^\dagger c, \\ 
H_{\rm int}/\hbar= & \, g_{mb} m^\dagger m q  - g_{cb}c^\dagger c q , \\ 
H_{d}/\hbar = &\, i \Omega\left ( m^\dagger e^{-i\omega_0t}-m e^{i\omega_0t} \right ) +iE\left ( c^\dagger e^{-i\omega_Lt}-ce^{i\omega_Lt} \right),
\end{split}
\label{Hamilt}
\end{equation}
are the free Hamiltonian, the interaction Hamiltonian, and the Hamiltonian of the drives of the magnon and cavity modes, respectively. $m$ and $c$ ($m^\dagger$ and $c^\dagger$) are the annihilation (creation) operators of the magnon and cavity modes, respectively, satisfying the commutation relation $[j,j^\dagger]=1$ ($j=m,c$). $q$ and $p$ ($[q,p]=i$) denote the dimensionless position and momentum of the deformation vibrational mode, modeled as a mechanical oscillator, which simultaneously couples to the magnon mode via the magnetostrictive interaction and to the cavity field via the radiation pressure. Both interactions are of the nonlinear dispersive type, and the corresponding bare coupling rates are $g_{mb}$ and $g_{cb}$. $\omega_m$, $\omega_b$ and $\omega_c$ are the resonant frequencies of the magnon, mechanical and cavity modes, respectively. The frequency of the magnon mode can be adjusted by varying the bias magnetic field $H_0$ via $\omega_m= \gamma H_0$, with the gyromagnetic ratio $\gamma/2\pi=28$ GHz/T for the YIG. In typical cavity magnonic experiments~\cite{S1,S2,S3}, $\omega_m/2\pi$ is about 10 GHz. The magnon mode is driven by a microwave magnetic field with amplitude $B_0$ and frequency $\omega_0$. The corresponding Rabi frequency $\Omega=\frac{\sqrt{5}}{4}\gamma\sqrt{N}B_0$ for a YIG sphere~\cite{JL18}, with $N$ the total number of spins. 
The cavity is weakly driven by a laser with frequency $\omega_L$ and power $P_L$, and the corresponding coupling strength $E=\sqrt{\kappa_cP_L/\hbar\omega_L}$, with $\kappa_c$ the cavity decay rate. 

By including the dissipation and input noise of each mode, and working in the interaction picture with respect to $\hbar\omega_0 m^\dagger m+\hbar\omega_L c^\dagger c$, the quantum Langevin equations (QLEs) of the whole system are given by
\begin{equation}\label{QLE1}
	\begin{split}
	 \dot{m}=&-\left ( i\Delta_m+\frac{\kappa_m}{2} \right )m-ig_{mb}mq+\Omega+\sqrt{\kappa_m}m^{\rm in} ,\\ 
	 \dot{c}=&-\left ( i\Delta_c+\frac{\kappa_c}{2} \right )c+ig_{cb}cq+E+\sqrt{\kappa_c}c^{\rm in} , \\
	 \dot{q}=& \omega_b p,\,\,\,  \dot{p}= -\omega_b q -\gamma_b p - g_{mb}m^\dagger m + g_{cb}c^\dagger c +\xi  ,
	\end{split}
\end{equation}
where $\Delta_m=\omega_m-\omega_0$ ($\Delta_c=\omega_c-\omega_L$) is the detuning of magnon (cavity) mode with respect to its drive field, and $\kappa_m$, $\kappa_c$ and $\gamma_b$ ($m^{\rm in}$, $c^{\rm in}$ and $\xi$) are the dissipation rates (input noises) of the magnon, cavity, and mechanical modes, respectively. The input noise operators have a zero expectation value and the following nonzero correlation functions~\cite{GC}: $\left \langle m^{\rm in}(t)m^{\rm in\dagger }(t') \right \rangle=\left [n_m(\omega_m)+1  \right ]\delta(t-t')$, $\left \langle m^{\rm in\dagger }(t)m^{\rm in}(t') \right \rangle=n_m(\omega_m)\delta(t-t')$, and $\left \langle c^{\rm in}(t)c^{\rm in\dagger }(t') \right \rangle=\delta(t-t')$, and a Markovian approximated $\delta$-correlated mechanical noise: $\left \langle \xi(t)\xi(t') +\xi(t')\xi(t)\right \rangle /2 \simeq \gamma_b\left [ 2n_b(\omega_b)+1 \right ]\delta(t-t')$, which is the case for a large mechanical quality factor $Q=\omega_b/\gamma_b\gg 1$~\cite{Kac}. The mean thermal excitation number $n_j(\omega_j)=1/[ \mathrm{exp}(\hbar\omega_j/k_BT)-1]$ ($j=m,b$) at an environmental temperature $T$.

Under continuous drives, the system will evolve to a steady state under the parameters constrained by the stability condition (which will be discussed in Sec.~\ref{noise}). When the amplitudes of the magnon and cavity modes are sufficiently large,  $\left | \left \langle m \right \rangle \right |,\, \left | \left \langle c \right \rangle \right |\gg1$, one can linearize the nonlinear dynamics around the steady-state averages by writing each mode operator as a classical average plus a quantum fluctuation operator, $k=\left \langle k \right \rangle+\delta k$ ($k=m,c,q,p$), and neglecting the small second-order fluctuation terms. As a result, the QLEs~\eqref{QLE1} are separated into two sets of linear equations for classical averages and quantum fluctuations, respectively. By solving the set of equations for classical averages, we obtain 
 \begin{equation}
 	\begin{split} 		
 		\left \langle m \right \rangle &= \frac{\Omega}{i\tilde{\Delta}_m + \frac{\kappa_m}{2}}, \,\,\,\,\,\,\,\,\,\,\,\,\, \left \langle c \right \rangle=\ \frac{E}{i\tilde{\Delta}_c + \frac{\kappa_c}{2}}, \\
 		\ \left \langle q \right \rangle =\,& \Big( g_{cb}\left | \left \langle c \right \rangle \right |^2-g_{mb}\left | \left \langle m \right \rangle \right |^2 \Big)/\omega_b, \,\,\,\,\,	 \left \langle p \right \rangle = 0 ,
 	\end{split}
 \end{equation}
where  $\tilde{\Delta}_m = \Delta_m + g_{mb}\left \langle q \right \rangle$ and $\tilde{\Delta}_c = \Delta_c - g_{cb}\left \langle q \right \rangle$ are the effective detunings. We consider a resonant laser drive, $\omega_L \simeq \omega_c$ ($\Delta_c \simeq 0$), which corresponds to an optimal situation for realizing high-sensitivity detection of the mechanical position by measuring the phase of the cavity field~\cite{JOSAB}. Under the resonant drive, the Stokes and anti-Stokes scattering probabilities are equal and the optomechanical interaction resembles a quantum nondemolition interaction~\cite{Grangier}. Furthermore, the cavity is weakly driven, leading to $g_{cb}\left | \left \langle c \right \rangle \right |^2 \ll g_{mb}\left | \left \langle m \right \rangle \right |^2$, such that the displacement is dominantly induced by magnetostriction, i.e., $\left \langle q \right \rangle \simeq  -\frac{g_{mb}}{\omega_b} \left | \left \langle m \right \rangle \right |^2$, which shows a linear dependance of $\left \langle q \right \rangle$ on the magnon population $N_m =|\left \langle m \right \rangle |^2$. This also ensures that the radiation pressure will yield a negligible backaction on the magnon population via the mediation of the mechanical oscillator.

\begin{figure}[b]
\hskip-0.12cm\includegraphics[width=0.5\linewidth]{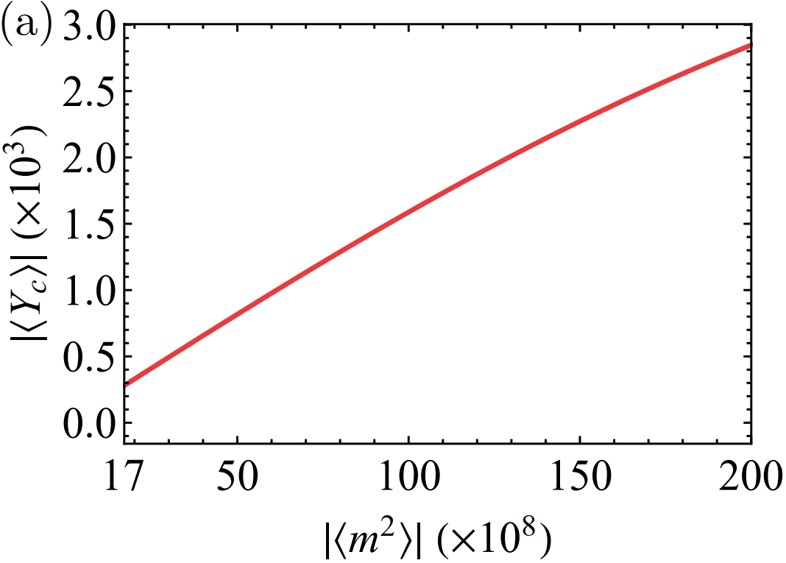} \vspace{15pt}
\includegraphics[width=0.5\linewidth]{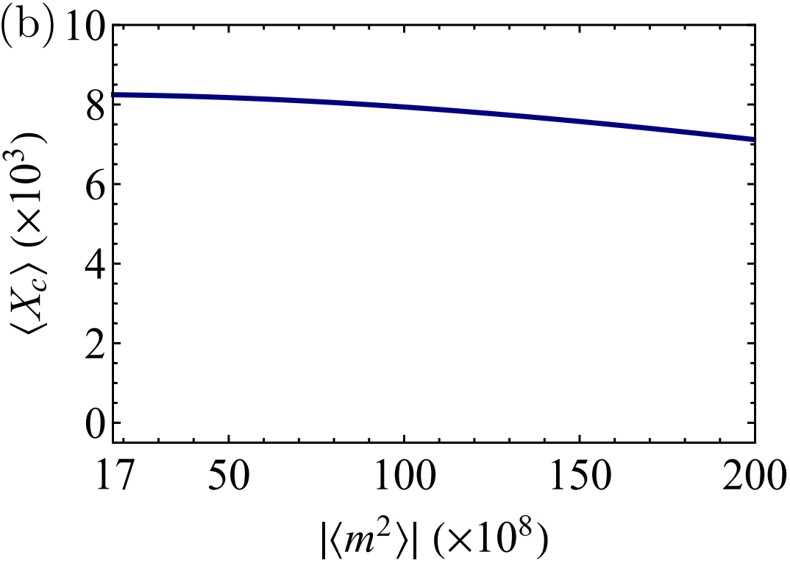}
\caption{Steady-state average of (a) the phase quadrature $\left | \left \langle Y_c \right \rangle \right |$ and (b) the amplitude quadrature  $ \left \langle X_c \right \rangle $ versus magnon population $\left | \left \langle m \right \rangle \right |^2$. See the text for other parameters.   }
\label{fig2}
\end{figure}

The tight connection between the mechanical displacement and the optical phase (under a resonant drive) offers the possibility to observe this linear dependance in the phase quadrature by homodyning the cavity output field (see Fig.~\ref{fig1}). This can be seen from the expressions of the phase and amplitude quadratures of the cavity field, given by  
\begin{equation}
\left \langle Y_c \right \rangle={-} \!\sqrt{2}\ \frac{E \tilde{\Delta}_c}{\tilde{\Delta}_c^2+\frac{\kappa_c^2}{4}} , \,\,\,\,\,\,  \left \langle X_c \right \rangle=\sqrt{2}\ \frac{ \frac{1}{2} E \kappa_c}{{\tilde{\Delta}_c}^2+\frac{\kappa_c^2}{4}}, 
\end{equation}
where $\tilde{\Delta}_c = -g_{cb}\left\langle q\right\rangle \simeq \frac{g_{cb} \, g_{mb}}{\omega_b} \left | \left \langle m \right \rangle \right |^2$, implying that the cavity frequency shift is mainly caused by the magnetostrictively induced displacement. Under the condition that this frequency shift is much smaller than the cavity linewidth $\tilde{\Delta}_c \ll \kappa_c$, we can approximately obtain  
\begin{equation}\label{YYY}
\left \langle Y_c \right \rangle \simeq {-}4\!\sqrt{2}\ \frac{E \tilde{\Delta}_c}{\kappa_c^2} \simeq  {-}4\!\sqrt{2} \, \frac{E g_{cb} g_{mb} }{\kappa_c^2 \omega_b} \left | \left \langle m \right \rangle \right |^2 , \\
\end{equation}
and a constant amplitude average $\left \langle X_c \right \rangle= \frac{ 2 \sqrt{2} E}{\kappa_c}$. Clearly, Eq.~\eqref{YYY} shows a linear dependence of the phase $\left \langle Y_c \right \rangle$ on the magnon population for a {\it fixed} laser power (more precisely, $E$). Though the condition $\tilde{\Delta}_c \ll \kappa_c$ leads to an excellent linear dependence, it places a limit on the maximum number of the magnon population that can be probed {\it linearly}. For a sufficiently large population (causing $\tilde{\Delta}_c$ to be comparable with $\kappa_c$), a nonlinear dependence starts to emerge in both $\left \langle Y_c \right \rangle$ and $\left \langle X_c \right \rangle$. This is seen in Figs.~\ref{fig2}(a) and~\ref{fig2}(b). Although in this nonlinear regime, one can measure both the phase and amplitude quadratures to determine the magnon population, we focus on the linear regime where one only needs to measure the phase quadrature, and the dependance is rather straightforward.  We have used the following parameters for Fig.~\ref{fig2}: $\omega_b/2\pi=10$ MHz, $g_{mb}/2\pi=1$ Hz, $\kappa_c/2\pi=100$ kHz, $g_{cb}/2\pi=10$ Hz, and a laser with power $P_L=1$ $\mu$W and wavelength $\lambda_L=1064$ nm.

\begin{figure}[t]
	\centering
\hskip-0.15cm	\includegraphics[width=0.48\linewidth]{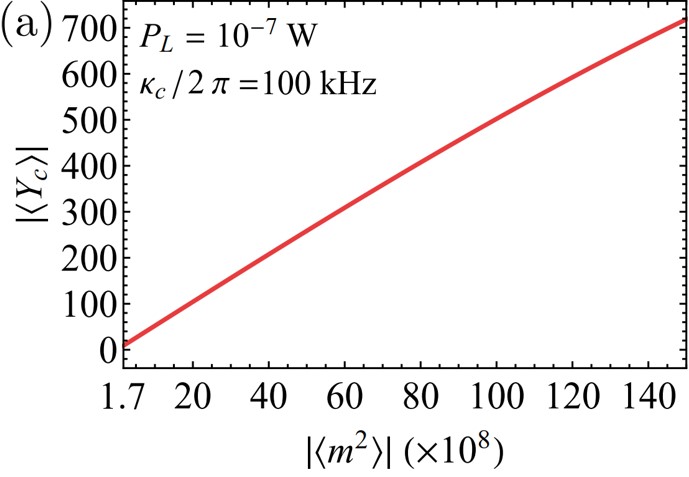}
	\includegraphics[width=0.48\linewidth]{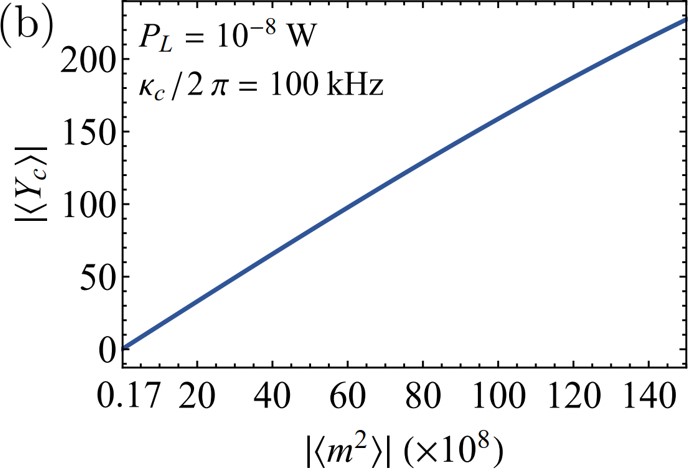}
		\vspace{6pt}
\hskip-0.15cm	\includegraphics[width=0.48\linewidth]{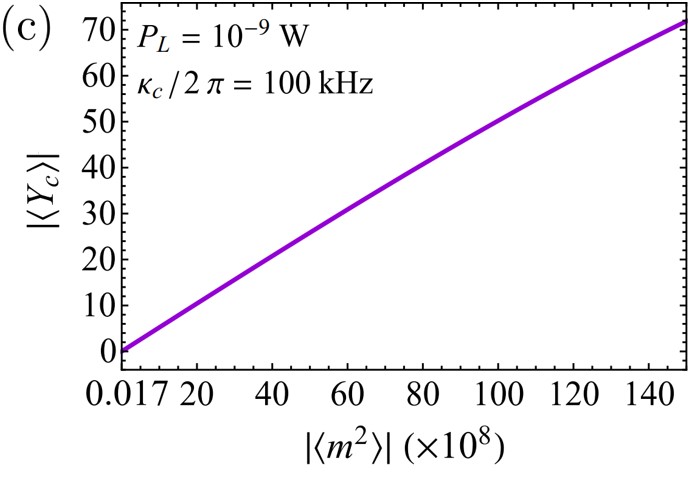}
	\includegraphics[width=0.48\linewidth]{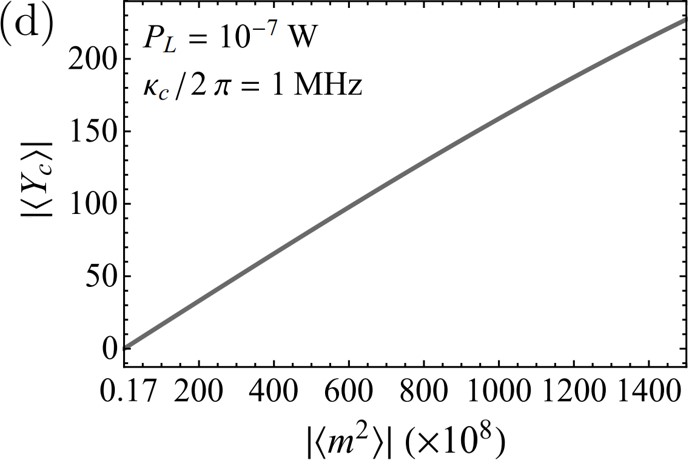}
\caption{Steady-state $\left | \left \langle Y_c \right \rangle \right |$ versus magnon population $\left | \left \langle m \right \rangle \right |^2$ at different laser powers: (a) and (d) $P_L=10^{-7}$ W, (b) $P_L=10^{-8}$ W, (c) $P_L=10^{-9}$ W. We take $\kappa_c/2\pi=100$ kHz in (a)-(c) and $\kappa_c/2\pi=1$ MHz in (d). The rest of the parameters are the same as in Fig.2. }
\label{fig3}
\end{figure}

Note that there is also a lower limit of measuring range of the magnon population determined by the perturbation condition $g_{cb}\left | \left \langle c \right \rangle \right |^2 \,{\ll}\, g_{mb}\left | \left \langle m \right \rangle \right |^2$ for a fixed laser power. In Figs.~\ref{fig3}(a)-\ref{fig3}(c), we show the phase average $\left |\left \langle Y_c \right \rangle \right|$ versus the magnon population for various laser powers. For each power, the measuring range is chosen to meet both the conditions for perturbation and the linear dependence. It shows that a larger power yields a larger $\left |\left \langle Y_c \right \rangle \right|$ (which increases the SNR, to be discussed in Sec.~\ref{noise}), but the measuring range is narrowed. In contrast, increasing $\kappa_c$ broadens the measuring range, but meantime reduces $\left |\left \langle Y_c \right \rangle \right|$ (thus the SNR), as revealed by comparing Figs.~\ref{fig3}(a) and~\ref{fig3}(d). So, there is a tradeoff between the SNR and the measuring range by changing $P_L$ or $\kappa_c$. The two conditions can be written in the following compatible form:
\begin{equation}
   \frac{g_{cb}^2}{\omega_b} \left | \left \langle c \right \rangle \right |^2    \ll   \frac{g_{cb} \, g_{mb}}{\omega_b} \left | \left \langle m \right \rangle \right |^2  \ll  \kappa_c,
\end{equation}
which sets an upper limit for the coupling strength $E \lll \kappa_c \sqrt{ \kappa_c \omega_b}/\big(2 g_{cb}\big)$, and thus for the laser power. This leads to an upper bound for $\left | \left \langle Y_c \right \rangle \right| / N_m$ from the result of Eq.~\eqref{YYY}, i.e.,
\begin{equation}\label{bound}
 \frac{ \left| \left \langle Y_c \right \rangle \right |} { N_m } \lll 2 \sqrt{2} \frac{ g_{mb}}{\sqrt{ \kappa_c \omega_b}}.
\end{equation}
It determines the highest possible SNR (for a given phase noise) and thus the resolution of magnons in our protocol. 
The upper bound suggests adopting a smaller cavity linewidth and a vibrational mode with a lower frequency but a larger magnon-phonon (dispersive) coupling.

\section{Effect of thermal noises and signal-to-noise ratio}\label{noise}

In the preceding section, we have established a linear dependence of the steady-state phase average on the magnon population. Although the phase average can be considered as the signal, there is also phase noise due to the presence of thermal noises from the environment. In particular, the phonon mode has a much lower frequency (typically in MHz for hundreds of microns sized YIG spheres~\cite{Tang,Davis}), and thus possesses the dominate thermal noise of the system. The corresponding SNR thus becomes a key indicator, which eventually determines the resolution of magnons of our approach. In what follows, we investigate the phase noise and the SNR in the steady state, and demonstrate that our protocol is still effective under substantial thermal noises.

The set of equations for quantum fluctuations can be written, in the quadrature form, as
\begin{equation}\label{QLEfluc}
\begin{split}
\delta \dot{X}_c&= - \frac{\kappa_c}{2} \delta X_c + \tilde{\Delta}_c  \delta Y_c  + \sqrt{\kappa_c} X_c^{\rm in},  \\
\delta \dot{Y}_c&= - \tilde{\Delta}_c \delta X_c  - \frac{\kappa_c}{2} \delta Y_c  + G_c \delta q  + \sqrt{\kappa_c} Y_c^{\rm in},  \\
\delta \dot{X}_m&= - \frac{\kappa_m}{2} \delta X_m + \tilde{\Delta}_m  \delta Y_m  + {\rm Im}G_m \delta q  + \sqrt{\kappa_m} X_m^{\rm in},  \\
\delta \dot{Y}_m&= - \tilde{\Delta}_m \delta X_m - \frac{\kappa_m}{2} \delta Y_m  - {\rm Re}G_m \delta q  + \sqrt{\kappa_m} Y_m^{\rm in},  \\
\delta \dot{q}&= \omega_b \delta p,   \\
\delta  \dot{p}&= - \omega_b \delta q \,{-} \gamma_b \delta p \,{-} {\rm Re}G_m \delta X_m \,{-} {\rm Im}G_m \delta Y_m \,{+} G_c \delta X_c \,{+} \xi,   \\
\end{split}
\end{equation}
where the quantum fluctuations of the quadratures are defined as $\delta X_k \,\,{=}\,\, \frac{1}{\sqrt{2}} \left ( \delta k+\delta k^\dagger \right )$, $\delta Y_k \,\,{=}\,\,\frac{i}{\sqrt{2}} \left ( \delta k^\dagger-\delta k \right )$, and the corresponding quadratures of input noises $X_k^{\rm in}= \frac{1}{\sqrt{2}} \left ( k^{\rm in} + k^{\rm in \dagger} \right )$, $Y_k^{\rm in}\,\,{=}\,\,\frac{i}{\sqrt{2}} \left ( k^{\rm in \dagger} - k^{\rm in} \right )$ ($k\,\,{=}\,\,c,m$). We consider a real optomechanical coupling $G_{c} \,{=}\,\sqrt{2}g_{cb}\left \langle c \right \rangle$, since $\left \langle c \right \rangle \,{\simeq}\, \left \langle c \right \rangle^*$ when $\tilde{\Delta}_c \,\,{\ll}\,\, \kappa_c$, and a complex magnomechanical coupling $G_{m} \,{=} \,\sqrt{2}g_{mb}\left \langle m \right \rangle$, as in general $\left \langle m \right \rangle \,{\neq}\, \left \langle m \right \rangle^*$. The equations~\eqref{QLEfluc} can be rewritten in a compact matrix form as
\begin{equation}
\dot{v}(t)={\cal A} v(t)+n(t),
\end{equation}
where $v(t)=\left [ \delta X_c(t), \, \delta Y_c(t), \, \delta X_m(t),\, \delta Y_m(t), \, \delta q(t) , \, \delta p(t) \right ]^T$, $n(t)=\left [\!\sqrt{\kappa_c}X^{\rm in}_c(t),\sqrt{\kappa_c}Y^{\rm in}_c(t),\sqrt{\kappa_m}X^{\rm in}_m(t),\sqrt{\kappa_m}Y^{\rm in}_m(t) ,0,\xi(t) \right ]^T$, and the drift matrix ${\cal A}$ is given by
\begin{equation}
	{\cal A}=\begin{pmatrix}
		-\frac{\kappa_c}{2} & \tilde{\Delta}_c & 0 & 0 & 0 & 0\\ 
		-\tilde{\Delta}_c & -\frac{\kappa_c}{2} & 0 & 0 & G_{c} & 0\\ 
		0 & 0 & -\frac{\kappa_m}{2} & \tilde{\Delta}_m & \textup{Im}G_{m} & 0\\ 
		0 & 0 & -\tilde{\Delta}_m & -\frac{\kappa_m}{2} & -\textup{Re}G_{m} & 0\\ 
		0 & 0 & 0 & 0 & 0 & \omega_b\\ 
		G_{c} & 0 & -\textup{Re}G_{m} & -\textup{Im}G_{m} & -\omega_b & -\gamma_b
	\end{pmatrix}.
\end{equation}
The system becomes stable when $t \to \infty$ if all the eigenvalues of the drift matrix ${\cal A}$ have negative real parts. This is equivalent to the stability condition obtained from the Routh-Hurwitz criterion~\cite{RH}, but the inequalities become quite involved for the present tripartite system. All of the results presented in this work satisfy this condition and are thus in the steady state.


The equations~\eqref{QLEfluc} can be solved conveniently in the frequency domain by taking the Fourier transform of each equation. After some algebra, we obtain the following solution for the phase fluctuation $\delta Y_c(\omega)$: 
\begin{equation}\label{Yc-noise}
\begin{split}
&\delta Y_c(\omega) = \chi_c(\omega) \,\chi_c^*(-\omega) \, \times \\ 
&\left[ -\tilde{\Delta}_c \! \sqrt{\kappa_c}  X_c^{\rm in}(\omega) + \left( \frac{\kappa_c}{2} {-} i \omega \right) \left( \!\! \sqrt{\kappa_c} Y_c^{\rm in}(\omega)  + G_c \chi_b(\omega) \Xi_{\rm back}(\omega)  \right)   \right ],
\end{split}
\end{equation}
where  $\Xi_{\rm back}(\omega)$ denotes the optomechanical backaction noise from the mechanical oscillator, given by
\begin{widetext}
\begin{equation}
\Xi_{\rm back}(\omega) = \frac{ \xi(\omega) + \! \sqrt{\kappa_c} \, \chi_c(\omega) \chi_c^*(-\omega) \, \CC (\omega) - \! \sqrt{\kappa_m} \, \chi_m(\omega) \chi_m^*(-\omega) \MM (\omega)  }{1- \chi_b(\omega) \left( G_c^2 \, \tilde{\Delta}_c \, \chi_c(\omega) \chi_c^*(-\omega) + \left| G_m \right|^2 \tilde{\Delta}_m \, \chi_m(\omega) \chi_m^*(-\omega)   \right)  },
\end{equation}
\end{widetext}
which further contains three noise sources: $i$) the noise directly from the mechanical thermal bath $\xi(\omega)$, $ii$) the optomechanical backaction noise from the cavity field $\CC (\omega)$, and $iii$) the magnomechanical backaction noise from the magnon mode $\MM (\omega)$, with
\begin{equation}
\begin{split}
 \CC (\omega) := \,& G_c \left[  \left( \frac{\kappa_c}{2} - i \omega \right) X_c^{\rm in}(\omega)  + \tilde{\Delta}_c Y_c^{\rm in}(\omega)   \right],  \\
 \MM (\omega) := \,& {\rm Re}G_m \left[ \left( \frac{\kappa_m}{2} - i \omega \right) X_m^{\rm in}(\omega)  + \tilde{\Delta}_m Y_m^{\rm in}(\omega)  \right]  \\
 &+  {\rm Im}G_m \left[  - \tilde{\Delta}_m X_m^{\rm in}(\omega)  +  \left( \frac{\kappa_m}{2} - i \omega \right) Y_m^{\rm in}(\omega)  \right].
 \end{split}
 \end{equation}
We have introduced the natural susceptibility $\chi_b (\omega)$ of the mechanical mode, $\chi_c (\omega)$ of the cavity field, and $\chi_m (\omega)$ of the magnon mode, given by
\begin{equation}
\begin{split}
\chi_b (\omega) &=  \frac{\omega_b}{\omega_b^2 - \omega^2 - i \gamma_b \omega},  \\
\chi_c (\omega) &=  \frac{1}{ \frac{\kappa_c}{2} + i (\tilde{\Delta}_c - \omega)} , \\ 
\chi_m (\omega) &=  \frac{1}{ \frac{\kappa_m}{2} + i (\tilde{\Delta}_m - \omega)}.
\end{split}
\end{equation}

\begin{figure}[b]
\hskip-0.6cm\includegraphics[width=0.9\linewidth]{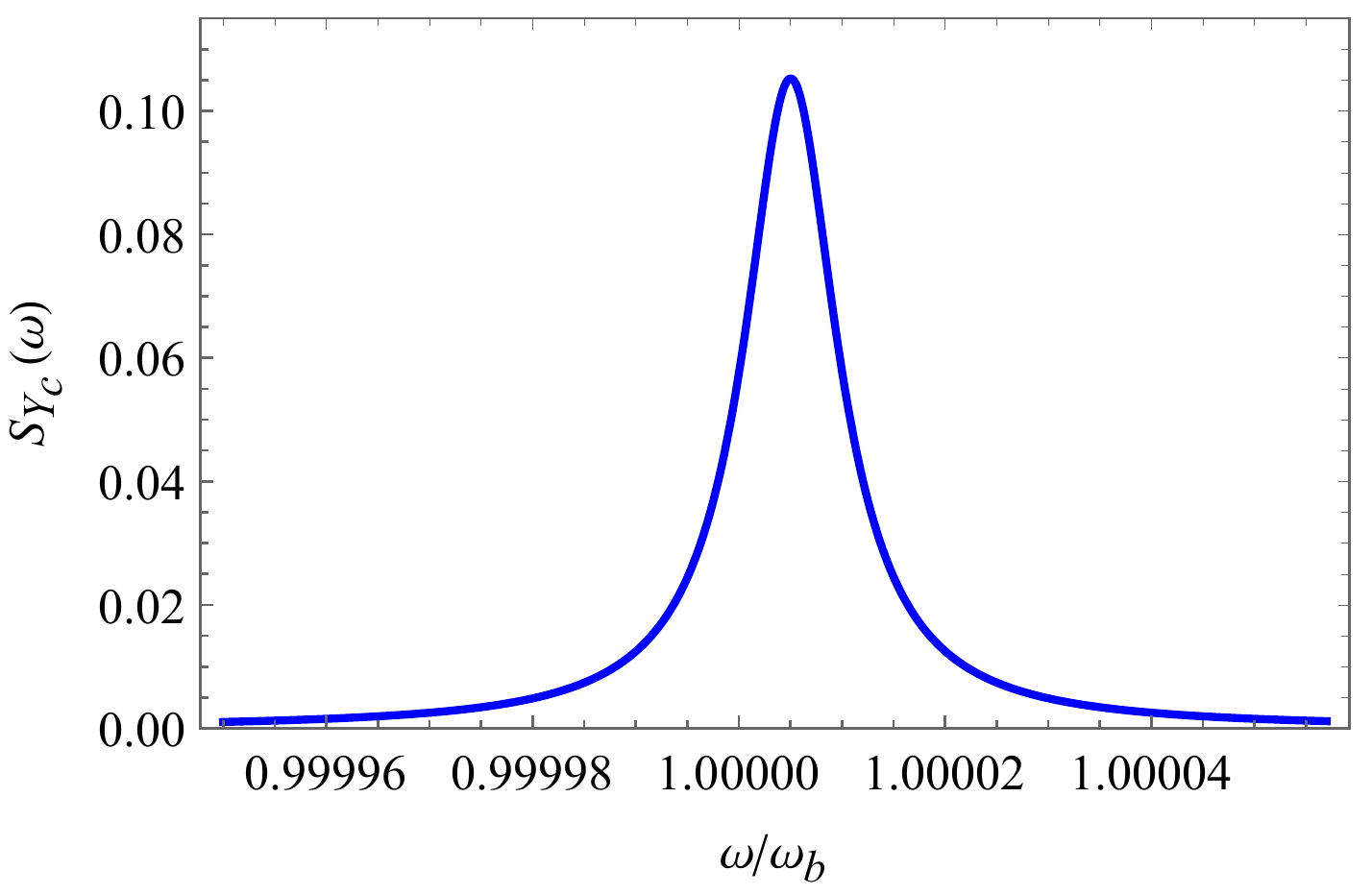}
\caption{NSD of the phase $S_{Y_c}(\omega)$ for $N_m=10^{10}$ and $T=293$ K. Other parameters are listed in the text.}
\label{fig4}
\end{figure}

\begin{figure}[t]
\hskip-0.36cm\includegraphics[width=0.8\linewidth]{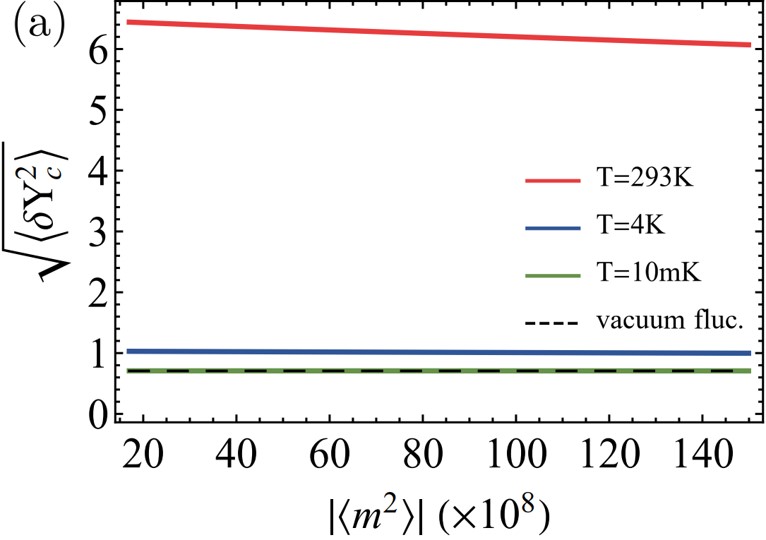}  \\  \vspace{9pt}
\hskip-0.3cm\includegraphics[width=0.794\linewidth]{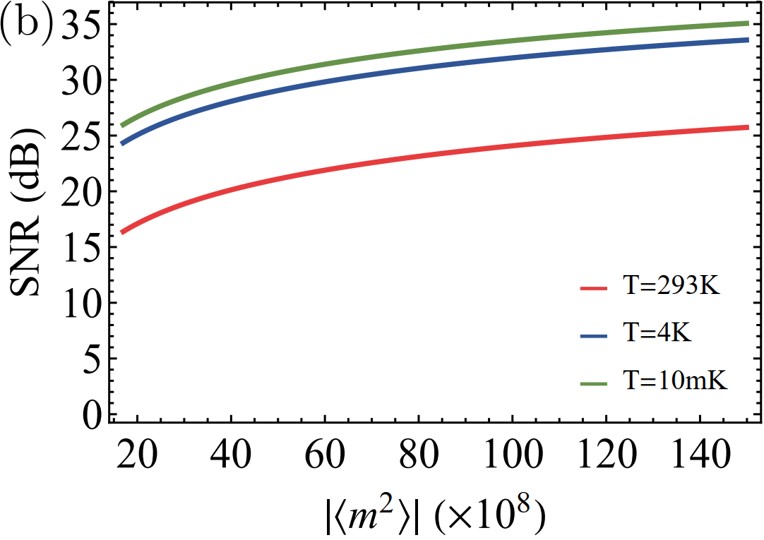}
\caption{(a) Standard deviation $\sqrt{\left \langle \delta Y_c^2 \right \rangle}$ and (b) SNR (in decibels) versus magnon population $\left | \left \langle m \right \rangle \right |^2$ at different temperatures $T=293$ K (red solid), $T=4$ K (blue solid), and $T=10$ mK (green solid). The black dashed line denotes vacuum fluctuations. The other parameters are those as in Fig.~\ref{fig4}, that is, the average $\left | \left \langle Y_c \right \rangle \right |$ used to calculate SNR is the same as in Fig.~\ref{fig2}(a).}
\label{fig5}
\end{figure}

From Eq.~\eqref{Yc-noise}, we can define the noise spectral density (NSD) of the phase quadrature
\begin{equation}
\begin{split}
S_{Y_c}(\omega) =  \frac{1}{4\pi} & \int_{-\infty}^{+\infty} \! d\omega' \, e^{-i (\omega+\omega') t } \,\,  \times   \\
 & \left \langle \delta Y_c (\omega) \delta Y_c  (\omega')  +  \delta Y_c  (\omega') \delta Y_c  (\omega) \right \rangle  .
\end{split}
\end{equation}
Here $S_{Y_c}(\omega)$ can be achieved using the input noise correlations in the frequency domain, which is shown in Fig.~\ref{fig4}. Clearly, a resonant peak appears at the mechanical frequency $\omega=\omega_b$. The parameters are those as in Fig.~\ref{fig2} and the others are $\omega_m/2\pi=10$ GHz, $\kappa_m/2\pi=1$ MHz, $\tilde{\Delta}_m = 0.5 \kappa_m$, $\gamma_b/2\pi=100$ Hz, $N_m=10^{10}$, and $T=293$ K.

The variance of the phase can be obtained by integrating $S_{Y_c}(\omega)$ over the frequency in the whole regime, i.e., 
\begin{equation}
\left \langle \delta Y_c(t)^2  \right\rangle = \frac{1}{2\pi} \! \int_{-\infty}^{+\infty} \!\! d\omega \, S_{Y_c}(\omega),
\end{equation}
which reflects the noise/imprecision of the phase.
In our notation, $\left\langle \delta Y_c(t)^2  \right\rangle = \frac{1}{2}$ denotes vacuum fluctuations, and the corresponding standard deviation $\sqrt{\left\langle \delta Y_c(t)^2  \right\rangle }\simeq 0.707$.

The classical average and the standard deviation of the phase allow us to define the SNR 
\begin{equation}
\textup{SNR}=\frac{\left | \left \langle Y_c \right \rangle \right |}{\sqrt{\left \langle \delta Y_c^2 \right \rangle}},
\end{equation}
 which is a key parameter and determines the resolution of magnons in our approach. The resolution can be defined as the value of the magnon population that can be detected with a unit SNR. It can be improved by enlarging the signal $\left | \left \langle Y_c \right \rangle \right |$, i.e., increasing $g_{mb}/\!\sqrt{ \kappa_c \omega_b}$ from Eq.~\eqref{bound}, or reducing the noise $\left \langle \delta Y_c^2 \right \rangle$. The latter can be realized by placing the system at cryogenic temperatures, or injecting a squeezed light to reduce the phase noise. In Fig.~\ref{fig5}(a), we show the phase noise (standard deviation) at $T=10$ mK, 4 K, and 293 K. Obviously, the phase noise is reduced by lowering the environmental temperature, which accordingly results in an improved SNR as displayed in Fig.~\ref{fig5}(b).  With the parameters employed in Fig.~\ref{fig5}, the SNR ranges from 24 to 35 dB at $T \le 4$ K (from 16 to 26 dB at room temperature) for $N_m \in [1.7\times 10^9,1.5\times 10^{10} ]$. The resolution of magnons is in the order of magnitude of $10^6$. The SNR can be further improved by using a squeezed light to break the standard quantum limit. A 15 dB squeezing~\cite{15dB} can improve the SNR by 7.5 dB, leading to the resolution in the order of magnitude of $10^5$. A potentially significant improvement is possible if design the system to have a much larger value of $ g_{mb}/\!\sqrt{ \kappa_c \omega_b}$, such as by increasing the cavity length $L$, since $\kappa_c \propto L^{-1}$, and reducing the sphere radius $R$ if using a YIG sphere, since $ g_{mb} \propto R^{-2}$ and $\omega_b \propto R^{-1}$~\cite{Tang}.

Lastly, it would be beneficial to compare our optical approach with the approaches using a superconducting qubit~\cite{Nak17,Nak20L}. The latter approaches can achieve very high resolutions of magnons. However, superconducting qubits require ultra-cold temperatures (tens of mK) to operate. Instead, our approach is hard to achieve as high resolution as in Refs.~\cite{Nak17,Nak20L} (however, depending on the system parameters), but our approach can work at room temperature and with a very high SNR. This would be highly appreciated for room-temperature experiments in cavity magnonics. Moreover, the methods of Refs.~\cite{Nak17,Nak20L} typically measure a very low magnon population. Our method can, however, measure a much wider range of the magnon population.

\section{Conclusion}\label{conc}

We propose an optical approach for measuring the steady-state magnon population in a ferromagnet or ferrimagnet. It utilizes the magnetoelasticity of the ferromagnet and the optomechanical coupling between the deformation displacement and an optical cavity. The linear dependence between the optical phase and the magnon population (under appropriate conditions) makes our protocol a good meter for measuring the steady-state magnon population. The study of the phase noise confirms that the protocol still works in the presence of thermal noises even at room temperature, reflected in its high SNR. We also provide strategies on how to improve the SNR and the resolution of magnon excitation numbers. We expect that our work can offer an alternative method for measuring the magnon population, as a complement to the existing approaches~\cite{prb21,Melkov,invSH1,invSH2,invSH3,BLS1,BLS2,Nak17,Nak20L}.

\section*{Acknowledgments}

This work has been supported by Zhejiang Province Program for Science and Technology (Grant No. 2020C01019), the National Natural Science Foundation of China (Grants Nos. U1801661, 11874249, 11934010, 12174329), and the Fundamental Research Funds for the Central Universities (No. 2021FZZX001-02).

\setcounter{figure}{0}
\renewcommand{\thefigure}{A\arabic{figure}}
\setcounter{equation}{0}
\renewcommand{\theequation}{A\arabic{equation}}

\section*{Appendix}

Here we provide the details on the quantization of the magnetization, the elastic strain, and the corresponding magnetoelastic energy, which lead to the Hamiltonian we use and a dominant dispersive coupling between magnons and phonons under appropriate assumptions.

The interaction between the magnetization and the elastic strain is described by the magnetoelastic coupling~\cite{Becker,Kittle}. In general, the magnetoelastic energy density is given by 
\begin{equation}
\begin{split}
	f_{me}=&\ \frac{b_1}{M_S^2}\left( M_x^2\epsilon_{xx} +M_y^2\epsilon_{yy}+M_z^2\epsilon_{zz}\right) \\
	&+\frac{2b_2}{M_S^2}\left( M_xM_y\epsilon_{xy}+M_xM_z\epsilon_{xz}+M_yM_z\epsilon_{yz}\right) ,
\end{split}\label{fme}
\end{equation}
where $b_1$ and $b_2$ are the magnetoelastic coupling coefficients (not specified), $M_S$ is the saturation magnetization, and $M_{x,y,z}$ are the corresponding magnetization components. The magnetoelastic energy density is related to the strain tensor $\epsilon_{ij}$, $\epsilon_{ij}=\frac{1}{2}\left ( \frac{\partial u_i}{\partial l_j}+\frac{\partial u_j}{\partial l_i}\right )$, where $u_i$ are the components of the displacement vector $\vec{u}$. 

The magnetization can be quantized (i.e. magnons) as
 \begin{equation}
  \hat{m}= \sqrt{\frac{V}{2\hbar\gamma M_S}}\left (M_x - i M_y \right),
 \end{equation}
 where $V$ is the volume of the ferromagnet. We therefore obtain
\begin{equation}
\begin{split}
	M_x=&\ \sqrt{\frac{\hbar\gamma M_S}{2V}}\left ( \hat{m}+\hat{m}^\dagger\right )\\
	M_y=&\ i\sqrt{\frac{\hbar\gamma M_S}{2V}}\left ( \hat{m}-\hat{m}^\dagger\right ),
\end{split}\label{Mxy}
\end{equation}
and consequently,
\begin{equation}
M_z= \left( M_S^2-M_x^2-M_y^2 \right)^{\frac{1}{2}} \approx M_S -\frac{\hbar\gamma }{V}\hat{m}^\dagger \hat{m}. 
\label{MMz}
\end{equation}

By replacing $M_{x,y,z}$ with Eqs.~\eqref{Mxy} and \eqref{MMz} in the magnetoelastic energy density $f_{me}$, and integrating over the whole volume of the ferromagnet, the first term in Eq.~\eqref{fme} yields the Hamiltonian (neglecting nonresonant fast-oscillating terms)
\begin{equation}
	H_1 \simeq \frac{b_1}{M_S}\frac{\hbar\gamma }{V}\hat{m}^\dagger \hat{m}\int dl^3\left ( \epsilon_{xx}+\epsilon_{yy}-2\epsilon_{zz}\right ), 
	\label{H111}
\end{equation}
which accounts for the dispersive interaction between magnons and phonons, and the second term leads to the Hamiltonian
\begin{equation}
\begin{split}
	H_2=& \,i \frac{b_2}{M_S}\frac{\hbar\gamma }{V}\left ( \hat{m}^2-\hat{m}^ {\dagger 2}\right )\int dl^3 \epsilon_{xy}  \\
	 + &\frac{2b_2}{M_S^2}\sqrt{\frac{\hbar\gamma M_S}{2V}}\left ( M_S-\frac{\hbar\gamma }{V}\hat{m}^\dagger \hat{m} \right )\left [\hat{m}\int dl^3\left ( \epsilon _{xz}+ i\epsilon _{yz}\right )+h.c. \right ],
\end{split}
\end{equation}
which describes the parametric magnon generation when the phonon frequency is twice the magnon frequency, or the linear magnon-phonon coupling when they are (nearly) resonant.

The present work utilizes the dispersive magnon-phonon coupling, corresponding to the situation where the phonon frequency is much lower than the magnon's. This typically occurs for a large-sized ferromagnet~\cite{Tang,Davis}.

The magnetoelastic displacement can be expressed as a superposition
\begin{equation}
	\vec{u}=\sum_{n,m,k} d^{\left( n,m,k\right) }\vec{\chi}^{(n,m,k)}\left(x,y,z \right), 
	\label{uvec}
\end{equation}
with $\vec{\chi}^{(n,m,k)}\left(x,y,z \right)$ the (normalized) displacement eigenmode and $d^{\left( n,m,k\right) }$ the corresponding amplitude. We can then quantize the mechanical motion as 
\begin{equation}
\begin{split}
 d^{\left( n,m,k\right)}= d_{\text{zpm}}^{\left( n,m,k\right)} \left(\hat{a}_{n,m,k}+ \hat{a}^\dagger_{n,m,k} \right) ,
 \label{dnmk}
\end{split}
\end{equation}
where $d_{\text{zpm}}^{\left( n,m,k\right) }$ is the amplitude of the zero-point motion, and $\hat{a}_{n,m,k}$ and $\hat{a}^\dagger_{n,m,k}$ are boson operators for each mode with specified mode indices $\left( n,m,k\right)$.  

Substituting Eqs.~\eqref{uvec} and \eqref{dnmk} into the dispersive interaction Hamiltonian $H_1$, we obtain
\begin{equation}
	H_1=\sum_{n,m,k}\hbar g_{mb}^{\left ( n,m,k\right )}\hat{m}^\dagger \hat{m} \left (\hat{a}_{n,m,k}+ \hat{a}^\dagger_{n,m,k}\right ),
\end{equation}
where $g_{mb}^{\left ( n,m,k\right )}$ is the magnon-phonon coupling strength, given by
\begin{equation}
	g_{mb}^{\left ( n,m,k\right )}=\frac{b_1}{M_S} \frac{\gamma}{V} \int dl^3  d_{\text{zpm}}^{\left( n,m,k\right) }\left ( \frac{\partial \chi_{x}^{(n,m,k)}}{\partial x}+\frac{\partial \chi_{y}^{(n,m,k)}}{\partial y}-2\frac{\partial \chi_{z}^{(n,m,k)}}{\partial z}\right ).
\end{equation}
If considering a specific mechanical mode and its motion in only one direction, we have the following Hamiltonian
\begin{equation}
	H_1^{\rm sing}=\hbar g_{mb} \hat{m}^\dagger \hat{m} \left (\hat{a}+ \hat{a}^\dagger\right ),
\end{equation}
that is the one we use in the Hamiltonian Eq.~\eqref{Hamilt} for the dispersive magnomechanical interaction (since $\hat{q}= \frac{1}{\sqrt{2}}  (\hat{a}+ \hat{a}^\dagger )$). For simplicity, we have omitted hat signs for the operators in the main text.


\begin{thebibliography}{99}
\bibitem{note}
The model is, however, not limited to a spherical structure and YIG, but valid for any ferrimagnet or ferromagnet with a dispersive magnon-phonon coupling and a detectable deformation displacement.
\end{thebibliography}
\end{document}